\documentclass[twocolumn, prb, superscriptaddress, showpacs]{revtex4-1}
\usepackage{amsmath, array}
\usepackage{graphicx}
\usepackage[dvipdfm]{hyperref}
\usepackage{xcolor}
\DeclareMathOperator{\tr}{Tr}
\newcommand{\exponential}[1]{\mathrm{e}^{#1}} 
\begin{document}

\title{Local environment effects
on a charge carrier injected into an Ising chain}

\author{Mirko M. M\" oller}
\affiliation{\!Department \!of \!Physics and Astronomy, \!University of\!
  British Columbia, \!Vancouver, British \!Columbia,\! Canada,\! V6T \!1Z1}
\author{Mona Berciu}
\affiliation{\!Department \!of \!Physics and Astronomy, \!University of\!
  British Columbia, \!Vancouver, British \!Columbia,\! Canada,\! V6T \!1Z1}
\affiliation{\!Quantum Matter \!Institute, \!University of British Columbia,
  \!Vancouver, British \!Columbia, \!Canada, \!V6T \!1Z4}

\date{\today}

\begin{abstract}
  We present numerically exact results for the spectral function of a single
  charge carrier that is injected into an Ising chain at finite temperature $T$.
  Both ferromagnetic and antiferromagnetic coupling between the Ising spins are
  considered. The interaction between the carrier and the Ising spins is assumed
  to be on-site and of Ising type, as well. We find that the carrier's spectral
  function exhibits a distinctive fine structure of resonances that are due to
  the temporary entrapping of the carrier inside small magnetic domains. The
  connection to models of disordered binary alloys where similar effects can
  occur is explained. We use these results to construct an accurate
  (quasi)analytic approximation for low and medium-$T$ spectral functions.
\end{abstract}
\maketitle

\section{Introduction}

Understanding the motion of charge carriers in magnetic backgrounds is
a problem that appears in a variety of materials under a variety of
circumstances. The magnetic order can be due to local or to itinerant
moments; their magnetic order may be intrinsic or may be mediated by
the charge carriers; the charge carriers may occupy the same or
different bands from those giving rise to the magnetic background; at
sufficiently low temperatures the background may order
ferromagnetically (FM), antiferromagnetically (AFM), in more
complicated patterns, or be frustrated; and so on and so forth.

One limit that has received a lot of attention is that of a single carrier
evolving in a magnetic background of local moments that have long-range magnetic
order in the absence of the carrier. If the carrier is injected into the band
that hosts the local  moments, its injection creates a hole in the magnetic
background. Hopping of the hole changes the magnetic background as it
corresponds to hopping in the opposite direction of a lattice spin.  This
situation is described by the so-called $t-J$ model which was first introduced by
Anderson \cite{Anderson} and then proposed as an effective model for cuprates by
Zhang and Rice \cite{Zhang-Rice, Dagotto}. Due to the possible connection to
cuprates, such models have received a lot of attention. 

This article, however, deals with a different case, where the carrier and the
magnetic moments are hosted by different bands. Such a situation occurs, for
instance, in FM semiconductors such as EuO and EuS and is described by the $s-f$
model \cite{Nolting-review}. In these materials the carrier is hosted in a wide
$s$-band, whereas the local moments are hosted by a partially filled, strongly
localized $f$-band. Here motion of the carrier does not necessarily modify the
configuration of the magnetic background.

For these latter models, if the lattice spins have  FM order at $T=0$, an exact
solution for the carrier's propagator exists. This is because the $z$-component
of the total spin (carrier plus local moments) is conserved so the number of
magnons that can be excited by the carrier is at most one (we assume that all
spins are ${1\over 2}$). Starting from the FM state either no spin flip is
possible (if the carrier has spin parallel to the background), or at most one
magnon can be created (if the carrier has spin antiparallel to the background).
The first case is trivial, and the second is simple enough to be solved
analytically. The solution was first discussed in Ref. \onlinecite{Shastry} and
then generalized to more complex lattices. \cite{Mona-sp} Analyses of the
eigenstates and eigenenergies are also available.\cite{Henning,Nakano}

For an AFM background, on the other hand, an analytic solution is not
known even at $T=0$ because the excitation of arbitrarily many magnons is
now possible as the carrier can exchange the spins of any background
pair with a spin-up and a spin-down. Combined with the generally very
nontrivial nature of the AFM background wavefunction, this complicates
matters exceedingly.

When trying to extend the FM solution to finite-$T$, one runs into
somewhat similar problems because the carrier is now injected into a
magnetic background which has a population of thermal magnons already
present. Approximations for this case have been proposed (i) for
medium to high-$T$, based on calculating moments of the Green's
function (GF) and fitting to limiting cases where the exact solution
is known. The temperature enters via the magnetization $m$, which is
assumed to be known and not to be affected by the carrier;
\cite{Nolting} (ii) for weak and strong interactions between the
carrier and the local moments, as well as fully FM ordered background
and paramagnetic background, Kubo used the coherent potential
approximation (CPA) to calculate the total density of states
(DOS).\cite{Kubo} CPA is a self-consistent approximation that assumes
that the carrier propagates in an effective medium. Furthermore, Kubo
uses the molecular field approximation for the interaction between
local moments, thereby ignoring correlations; (iii) for very low-$T$,
an approximation for the self-energy was obtained by us by limiting
the states that enter the thermal average to one-magnon
states.\cite{low-T} It shows unphysical features at higher
temperatures where higher energy states become relevant.

None of these approaches includes fully the effects of the local
environment on the carrier. In this article we try to answer the
question of what role, if any, is played by the local environment. Is
there a class of configurations that contributes more than others? How
does an AFM ordered environment compare to a FM ordered one, and are
there any similarities between the two? Answering such questions is
crucial for the development of a better understanding (and hopefully
better approximations) for the finite-$T$ spectral weight of the
carrier.

Our approach is to study a simplified model which allows us to obtain the first
(to the best of our knowledge) finite-$T$ exact numerical results in the
thermodynamic limit. These teach us valuable lessons about the physics of this
simpler problem; some of these are relevant for more complex models, too. The
simplification consists in neglecting all spin-flip processes, i.e. assuming
that all interactions between the lattice spins, and between them and the
carrier, are Ising-like. While the $T=0$ behavior of this model is trivial, we
find that it produces rich physics at finite-$T$, which needs to be understood
before considering the additional complications introduced by allowing spin-flip
processes.

Apart from making an exact solution possible, this simple model also
allows us to consider both FM and AFM backgrounds on equal footing,
which is not possible in general. Interestingly, we find considerable
similarities between the results for the two cases, which we are able
to explain as being due to similarities between the local background
configurations that control the (at least low and medium-$T$) carrier
dynamics. We then use these insights to construct an analytical
approximation which accurately reproduces the features of the exact
solution in the range of low to medium temperatures.

By ignoring spin-flip processes, we create a situation that is in some
ways similar to that of disorder binary alloys of the type
A$_x$B$_{1-x}$. A carrier in the conduction band of such an alloy
experiences a different on-site energy if it sits on an A or a B
site. In our model the on-site carrier energy is $\pm J_0$, depending
on whether the local spin at that site points up or down. Thus, any
configurations of the alloy can be exactly mapped into a spin
configuration of the magnetic background.

These binary alloys have been extensively studied. An exact numerical
solution can be obtained for the disorder-averaged GFs, as pointed out
by Schmidt\cite{Schmidt} and Dyson.\cite{Dyson} It was used for
numerical studies of the phonon DOS \cite{Dean} and the DOS of
carriers in the conduction band.\cite{Alben, Economou} Many
approximations have been proposed, the most well-known and used being
the CPA.\cite{Soven} Diagrammatic expansions of the self-energy have
been developed by several authors, see for instance
Refs. \onlinecite{Yonezawa, Nickel1, Butler}, but were shown to
result in non-analytic behavior and thus unphysical GFs.\cite{Nickel2} Later
this problem was resolved by a careful consideration of which diagrams
to sum.\cite{Mills}

This knowledge cannot be directly used in our problem, despite the
similarities between the models, because for binary alloys one
performs a disorder average whereas we do a thermal average. The
disorder average keeps the concentration $x$ fixed; all configurations
consistent with it are equally likely (one generally ignores
correlations in the disorder), all the others are forbidden. In
contrast, a thermal average includes all possible configurations but
with a Boltzmann factor that controls the extent of spin-spin
correlations. These correlations are key, as no magnetic order can
exist in their absence. The problems also have different symmetry
properties. Our problem is truly translationally invariant, whereas
for binary alloys translational invariance is only restored by
averaging over all possible disorder configurations. For example,
while all eigenstates are localized in a 1D disordered alloy model,
the eigenstates of our model remain extended at any $T$. Despite these
differences, there are similarities between the resulting spectral
weights which help us understand our results, as discussed below.

This article is organized as follows. Section II presents the model
and Section III describes the method by which we calculate the GF. 
Results are discussed in Section IV for a 1D chain, although the
method can be trivially extended to higher dimension. We also present
here the analytic approximation we propose for low and medium temperatures.
The concluding remarks are in Section V.

\section{Model}

Consider a single spin-${1\over 2}$ charge carrier which interacts
with a chain of Ising spins, also of magnitude ${1\over 2}$. Note that
the exact method that we use can be generalized straightforwardly to
higher dimensions. The advantage of 1D chains (apart from speed of
computations) is that a host of analytic results are available for
the undoped case. We therefore limit ourselves here to Ising chains
with periodic boundary conditions after $N\rightarrow \infty$ sites.
The $n^\text{th}$ site is located at $R_n=n a$, and we set $a=1$.

The exchange between lattice spins is Ising-like:
\begin{align}
  \hat{H}_{\text{I}} = -J \sum_i \hat{\sigma}_i \hat{\sigma}_{i+1}
  -h\sum_i \hat{\sigma}_i,
\end{align}
and is FM ($J>0$) or AFM ($J<0$). The second term describes the effect
of an external magnetic field $h$. The spin operator at site $i$ is
$S_i^{z} = {1\over 2} \hat{\sigma}_i$, with the prefactors absorbed
into the coupling constants. The eigenstates $ \hat{H}_{\text{I}}
|\{\sigma \}\rangle =E_I(\{\sigma \})|\{\sigma \}\rangle$ are
described by the set $\{\sigma\}\equiv
\{\sigma_1,\sigma_2,\dots,\sigma_N\}$ of eigenvalues $\sigma_i=\pm 1$
of each spin, and $E_I(\{\sigma\})=-J\sum_i \sigma_i \sigma_{i+1} - h
\sum_i \sigma_i$ is the eigenenergy.

The kinetic energy of the carrier is described by a tight-binding
model
\begin{align}
  \hat{T}=-t\sum_{i,\sigma} \left(c_{i,\sigma}^\dagger c_{i+1,\sigma}
  + \text{h.c.}\right)= \sum_{k,\sigma}^{}\varepsilon(k)c_{k,\sigma}^\dagger c_{k,\sigma},
\end{align}
where $c_{i,\sigma}^\dagger$ is the creation operator of a
spin-$\sigma$ carrier at site $i$. The creation operator in
momentum-space is $c_{k,\sigma}^\dagger = {1\over \sqrt{N}}\sum_n
\exponential{i k R_n} c_{n,\sigma}^\dagger$, and $\varepsilon(k)= -2 t \cos(k)$.

The exchange between the carrier and the lattice spins is also of
Ising type:
\begin{align}
  \hat{H}_{\text{exc}} = J_0 \sum_{i,\sigma} \sigma c_{i,\sigma}^\dagger
  c_{i,\sigma} \hat{\sigma}_i,
\end{align}
where again a factor $1/4$ is absorbed into $J_0$. Since no spin-flips
are allowed in this model, from now on we assume without loss of
generality that the carrier has spin-up, $\sigma=\uparrow$, and not
write it explicitly. We also set $J_0>0$. Results for a spin-down
carrier are obtained from these by switching 
$J_0\rightarrow - J_0$.

For the undoped Ising chain, an exact solution at
finite-$T$ is possible and reviewed in the appendix. Note that a 1D chain
has long-range magnetic order only at $T=0$. We can,
however, mimic a finite-$T$ ordered state by turning on the magnetic field
$h$. 
This leads to a finite correlation independent of the distance between
spins.

\section{Method}

We want to calculate the finite-$T$ retarded single-particle GF, which
in the time-domain is defined as:
\begin{align}
  G(k,\tau) = -i \theta(\tau) \left \langle c_k(\tau) c_k^\dagger(0)
  \right \rangle_T
  \label{eq:GFtime}
\end{align}
where the operator $c_k(\tau) = \exponential{i \hat{H} \tau} c_k \exponential{-i
\hat{H} \tau}$ is in the Heisenberg picture.  Furthermore $\left \langle \dots
\right \rangle_T= \frac{1}{Z}\sum_{\{\sigma\}} \langle \{\sigma\} |
\exponential{-\beta E_I(\{\sigma\})}\dots | \{\sigma\} \rangle$ is the thermal
average and $\beta = 1/(k_B T)$.  From now on we set $\hbar=1, k_B=1$.

It needs to be pointed out that the temperature dependence in Eq.
(\ref{eq:GFtime}) is solely controlled by $\hat{H}_{\text{I}}$: we
inject the carrier into the otherwise undoped Ising chain, which at a
given $T$ is described by a density matrix
$\hat{\rho}_I=\exponential{-\beta \hat{H}_{\text{I}}}/Z$. In other
words, this formulation is for a canonical ensemble with a conserved
carrier number, not the usual grand-canonical formulation.

It is convenient to work in the frequency-domain. We Fourier
transform $G(k,\omega) = \int_{\infty}^{-\infty} \mathrm{d} \tau
\ \exponential{i \omega \tau} G(k,\tau)$\ which yields:
\begin{align}
  G(k,\omega) = \left \langle c_k \hat{G}(\omega+\hat{H}_I)
  c_k^\dagger \right \rangle_T,
  \label{eq:GFomega}
\end{align}
where $\hat{G}(\omega) = (\omega-\hat{H}+i \eta)^{-1}$\ is the
resolvent of $\hat{H}$. The small, real quantity $\eta>0$ ensures
retardation and sets a finite carrier lifetime $1/\eta$. Note that
the argument of the resolvent in Eq. (\ref{eq:GFomega}) is shifted by
$\hat{H}_I$. This means that the energy is measured from that of the
Ising chain at the time of injection. This becomes clear when using a
Lehmann representation by projecting on the one-carrier eigenstates
$\hat{H}| \psi_n \rangle = E_n| \psi_n \rangle $:
\begin{align}
  G(k, \omega) = \sum_n \sum_{\{\sigma\}} \frac{\exponential{-\beta
      E_I(\{\sigma\})}}{Z} \frac{\langle \{\sigma\} | c_k | \psi_n
    \rangle \langle \psi_n | c_k^\dagger | \{\sigma\} \rangle }{\omega
    + E_I(\{\sigma\}) - E_n + i \eta}
  \label{eq:Lehmann}
\end{align}
$G(k,\omega)$ has poles at energies $\omega=E_n-E_I(\{\sigma\})$ that
measure the change in total energy due to the carrier's injection. The
weights correspond to the overlap between the true eigenstates $|
\psi_n \rangle$ and the free-carrier states $c_k^\dagger | \{\sigma\}
\rangle$.

For any configuration $\{\sigma\}$, the contribution to the
thermal average can be evaluated using continued
fractions.\cite{ConFrac} First, we need to shift to the real-space
representation. Making use of the translational invariance we then
obtain
\begin{align}
  G(k,\omega) = \sum_{n} \exponential{i k R_n} \left \langle
  g_{0,n}(\omega, \{\sigma\}) \right \rangle_T,
  \label{eq:GFRealSpace}
\end{align}
where we define:
\begin{align}
  g_{m,n}(\omega, \{\sigma\}) = \langle \{\sigma\} | c_m
  \hat{G}(\omega+\hat{H}_I) c_n^\dagger | \{\sigma\} \rangle .
\end{align}
Physically $g_{m,n}(\omega, \{\sigma\})$ is related to the probability
that the carrier is injected at site $n$ and propagates to site
$m$. For any state $\{\sigma\}$ (except the fully ordered ones) the
translational invariance is broken, $g_{m,n}(\omega, \{\sigma\})\ne
g_{0,n-m}(\omega, \{\sigma\})$. This symmetry is only restored by the
ensemble average, which then leads to Eq. (\ref{eq:GFRealSpace}).

To obtain the GF equations of motion (EOM) we use Dyson's identity
$\hat{G}(\omega)=\hat{G}_0(\omega)+\hat{G}(\omega)\hat{V}\hat{G}_0(\omega)$
where $\hat{H}=\hat{H}_0+\hat{V}$ and 
 $\hat{G}_0(\omega)$ is the resolvent for $\hat{H}_0$.
Choosing $\hat{H}_0=\hat{H}_I + \hat{H}_{\text{exc}}$, we find:
\begin{align}
  g_{0,n}(\omega, \{\sigma\}) = &G_0(\omega-J_0 \sigma_n)
  \left(\delta_{0,n}-t g_{0,n-1}(\omega, \{\sigma\}) \nonumber \right . \\
  & \left . -tg_{0,n+1}(\omega,\{\sigma\}) \right),
\end{align}
where $G_0(\omega)=(\omega+i \eta)^{-1}$. Since the EOM do not change
the values of $\omega$ and $\{\sigma\}$, to shorten notation we do not write
them explicitly in the following.

The EOM are solved with the ansatz \cite{ConFrac} $g_{0,n} = A_n
g_{0,n-1}$ for $n > 0$ and $g_{0,n} = B_{-n} g_{0,n+1}$ for $n<0$
($N\rightarrow \infty$ is assumed and implemented as explained
below). Then
\begin{align}
  A_n = \frac{-t G_0(\omega-J_0 \sigma_n)}{1+tG_0(\omega-J_0
    \sigma_n)A_{n+1}},
\end{align}
and similarly for $B_n$. We now introduce a cutoff $M_c\gg 1$ at
which we truncate these relations by setting
$A_{M_c+1}=B_{M_c+1}=0$. The justification is provided by the finite
lifetime $1/\eta$ of the carrier, which prevents it from propagating
arbitrarily far from its injection site. As a result $g_{0,n}$, which
measures the amplitude of probability that the carrier propagates
between sites $n$ and 0, must vanish for sufficiently large
$|n|$.\cite{ConFrac}

It is then straightforward to calculate all $A_1, \dots , A_{M_c}$ and
$B_1, \dots, B_{M_c}$ for the configuration $\{\sigma\}$\ and a given
$\omega$, to find:
\begin{align}
& g_{0,0}= \frac{1}{\omega - J_0 \sigma_0 +t(A_1+B_1)}\\ & g_{0,n} =
  A_{M_c} \dots A_1 g_{0,0}, \mbox{ if } n>0 \\ & g_{0,-n} = B_{M_c}
  \dots B_1 g_{0,0}, \mbox{ if } n<0 .
\end{align}

Let us now discuss the cutoff $M_c$ in more detail. In practice $M_c$
is increased until convergence is reached. Since $g_{0,0}$ for
the fully ordered configuration is known analytically, it can be used
to verify the convergence.

We need to have $M_c \ll N/2$, otherwise the carrier may travel
between sites $n$ and 0 on both sides of the closed loop, which is at
odds with the ansatz chosen above. This condition is automatically
satisfied if $N \rightarrow \infty$. To take this limit, we note that
$g_{0,n}$ only depend on the spins $\sigma_{-M_c}, \sigma_{-M_c+1},
\ \dots\ , \sigma_{M_c}$. We make use of this by splitting the full
set $\{\sigma\}$ into the set $\{M_c\}$\ containing just the
aforementioned spins, and the complementary set $\{M_c\}^C$. The
energy of the Ising chain is also split: $E_I(\{\sigma\})=
E_I(\{M_c\})+ E_I(\{M_c\}^C, \sigma_{-M_c},\sigma_{M_c} )$, where
\begin{align}
E_I(\{M_c\}) = -J\sum_{n=-M_c}^{M_c-1} \sigma_{n} \sigma_{n+1}
-h\sum_{n=-M_c}^{M_c} \sigma_n
\end{align}
and $E_I(\{M_c\}^C, \sigma_{-Mc}, \sigma_{M_c})$ contains the
energy of all other bonds and spins, including the ``boundary'' bonds
$\sigma_{M_c}\sigma_{M_c+1}$ and $\sigma_{-M_c-1} \sigma_{-M_c}$. This
is why it also depends on $\sigma_{\pm M_c}$, not just on the
$\{M_c\}^C$ spins.

Eq. (\ref{eq:GFRealSpace}) can then be rewritten as:
\begin{align}
  G(k,\omega) = &\sum_n \exponential{i k R_n} \sum_{\{M_c\}}
  \frac{\exponential{-\beta E_I(\{M_c\})} }{Z} g_{0,n}(\omega,
  \{M_c\}) \nonumber \\ &\times
  Z_{\text{bath}}(\beta,\sigma_{-M_c},\sigma_{M_c}),
  \label{eq:Zbath}
\end{align}
where
\begin{align}
  &Z_{\text{bath}}(\beta,\sigma_{-M_c},\sigma_{M_c}) =
  \sum_{\{M_c\}^C} \exponential{-\beta E_I(\{M_c\}^C,\sigma_{-Mc},
    \sigma_{M_c})}.
\end{align}
$Z_{\text{bath}}$\ is the partition function of the complementary set
of spins $\{M_c\}^C$, for set values of its ``boundary'' spins
$\sigma_{-M_c}$ and $\sigma_{M_c}$. Using transfer matrices (see Appendix for details), we find
that:
\begin{align}
  Z_{\text{bath}}(\beta,\sigma_{M_c},\sigma_{-M_c}) =
  \left(\mathcal{T}^{N-2M_c}\right)_{\sigma_{M_c},\sigma_{-M_c}}.
\end{align}
Thus, $\lim_{N\rightarrow \infty}
Z_{\text{bath}}(\beta,\sigma_{M_c},\sigma_{-M_c})/Z$ is known analytically. The
average in Eq. (\ref{eq:Zbath}) now involves only the spins $\{M_c\}$ in the
chain sector that can be explored by the carrier within its finite lifetime
$1/\eta$. Effectively, the rest of the infinite chain is treated as a bath that
is integrated out analytically. We use the Metropolis algorithm \footnote{The
autocorrelation time ranges from 4 Monte Carlo (MC) steps ($\beta t = 1$) to
6630 MC steps ($\beta t = 5$). During one MC steps an attempt to flip each of
the $2M_c+1$\ Ising spins is made exactly once. For most spectra 204800 MC
measurements were used.  The time waited between measurements is identical to
the autocorrelation time.} to estimate the sum over the $\{M_c\}$ set. The
results are discussed next.

\section{Results}
\subsection{T=0 result}
We begin by briefly reviewing the $T=0$ solution which can be
calculated exactly and serves as a useful reference.

For FM coupling $J>0$, at $T=0$ all Ising spins point either up or
down, $m=\pm1$. Then $\hat{H}_{\text{exc}}$ simply shifts the energy
of the carrier by $J_0 m$: $E_m^{\text{FM}}(k) = \varepsilon_k +J_0 m$.
As $T\rightarrow 0$ and for $h=0$, an infinite chain will arbitrarily
choose as its ground state one of these two possible FM
configurations. One can control which configuration is chosen by
cooling the system in a small magnetic field, which is then 
switched off. However, as long as the temperature is not exactly zero
and if $h=0$, then the presence of large domains with opposite order is
possible, especially in 1D. If the carrier is injected into one of
these large domains it is unable to leave it within its finite
lifetime $1/\eta$. Consequently as $T\rightarrow 0$ and for $h=0$ we expect to
see contributions from both subspaces $m<0$ and $m>0$, and the GF
becomes:
\begin{align}
  G^{\text{FM}}(k,\omega) = \frac{1}{2} \left
  (\frac{1}{\omega-E^{\text{FM}}_++i \eta} + \frac{1}{\omega
    -E^{\text{FM}}_-+i \eta} \right ).
\end{align}

For AFM coupling there are also two possible ground states: either all spins of
the even sublattice point up and all spins of the odd sublattice point down, or
vice versa. The doubling of the unit cell results in the appearance of two bands
in the reduced Brillouin zone $(-\pi/2,\pi/2]$, with energies
$E^{\text{AFM}}_{\pm}(k) = \pm \sqrt{J_0^2 + \varepsilon_k^2}$. Averaging over
both contributions for the reasons discussed above, the $T\rightarrow 0, h=0$ GF
is found to be:
\begin{align}
  G^{\text{AFM}}(k,\omega) = \frac{\omega + \varepsilon_k + i \eta}
  {(\omega-E^{\text{AFM}}_+ + i \eta)(\omega-E^{\text{AFM}}_- + i
    \eta)}
\end{align}
for any $k\in (-\pi, \pi]$.

\begin{figure}[t]
 \includegraphics[width=\columnwidth, clip]{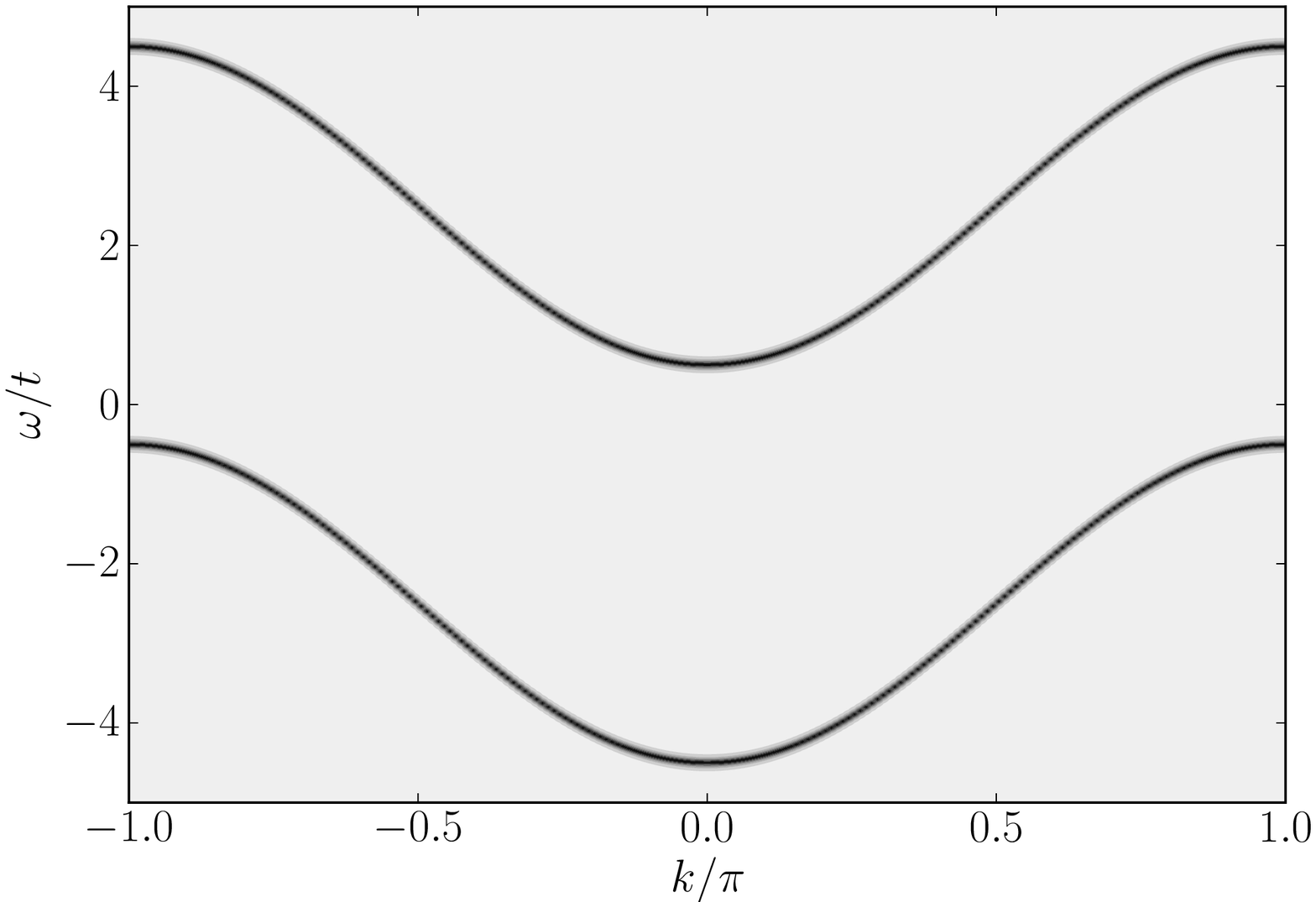}
 \includegraphics[width=\columnwidth, clip]{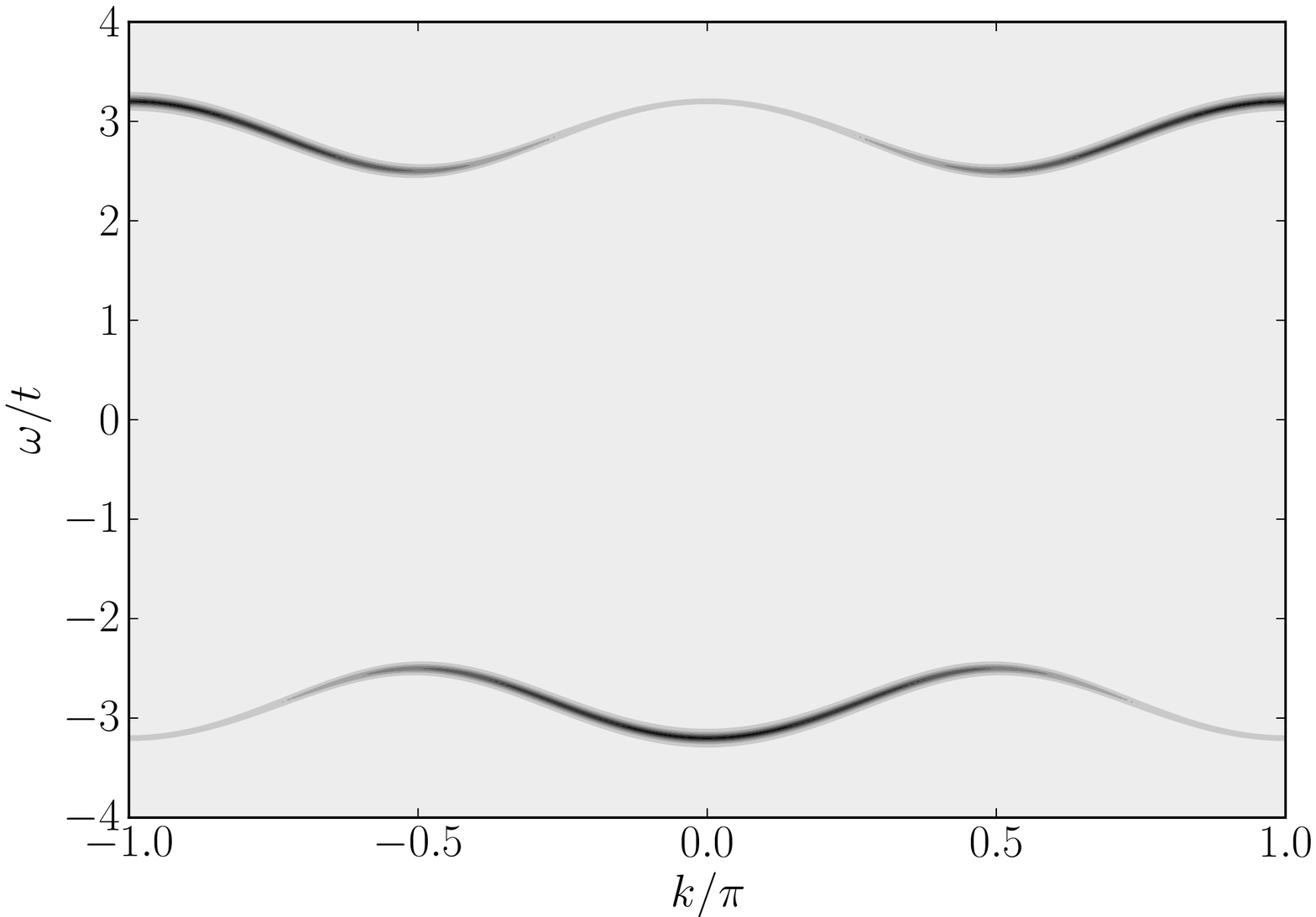}
  \caption{Contour plots of the $T=0$ spectral functions for FM (top) and AFM
  (bottom) coupling, for $|J|/t=0.5,\ J_0/t=2.5,\ h=0,\ \eta / t=0.04$.}
  \label{fig:ZeroTFM}
\end{figure}

Contour plots of the spectral function $A(k,\omega) = -\frac{1}{\pi}
\text{Im} G(k,\omega)$ for these GFs are shown in
Fig. \ref{fig:ZeroTFM}. As expected from applying the Lehmann
representation, Eq. (\ref{eq:Lehmann}), to these GFs, the spectrum
consists of two bands for both FM and AFM coupling. For FM coupling,
the bands have bandwidths of $4t$, are centered at $\pm J_0$ and have
equal weights for all $k$. For AFM coupling, the bands span
$[-\sqrt{J_0^2+4t^2}, -J_0]$ and $[J_0, \sqrt{J_0^2+4t^2}]$,
respectively. The eigenenergies show the $\pi$ periodicity expected
for the two-site unit cell. Spectral weight is transferred from the
lower to the upper band as $|k|$ increases, because the GF combines
contributions from both sublattices with a $k$-dependent phase factor.

Note that for $J_0/t \leq 2$ the two FM bands overlap. In order to
simplify future analysis, we set $J_0/t=2.5$ from now on. We also set
$|J|/t=0.5$, although we note that if $h=0$, $J$ only appears in
conjunction with $\beta$ so a choice for $J$ simply sets the
temperature scale. We use a cutoff of $M_c=400$, which for
$\eta$=0.04 is sufficient for convergence for the fully ordered FM
chain (this is the most slowly converging case).

\begin{figure}[t]
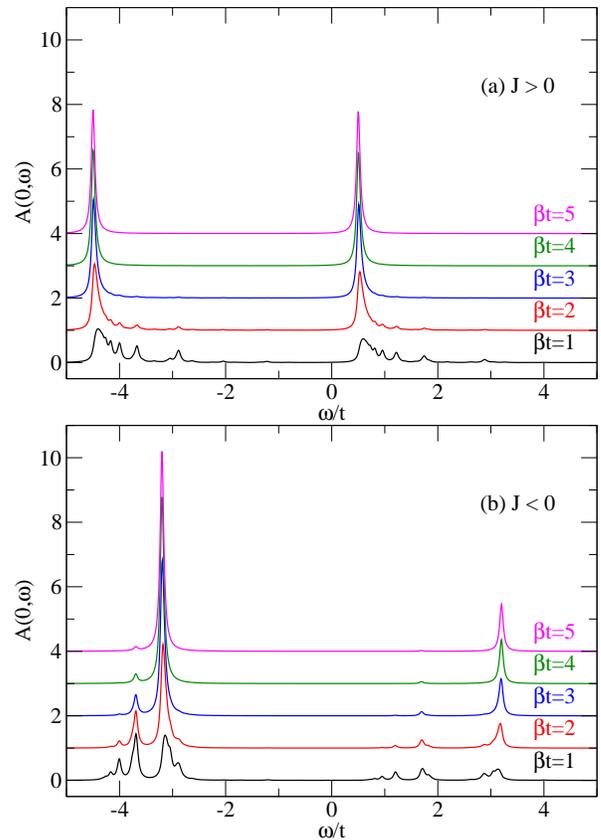

  \includegraphics[width=0.9\columnwidth]{fig2a.eps}
  \includegraphics[width=0.9\columnwidth]{fig2b.eps}
  \caption{(color online) Spectral function $A(0,\omega)$ for different
  temperatures and FM (top) and AFM (bottom) coupling. Parameters are
  $|J|/t=0.5, \ J_0/t=2.5, \ h=0,\ \eta / t=0.04$.}
  \label{fig:DiffBetaFM}
\end{figure}

\subsection{finite-T numerical results}
We now discuss the results of the Monte Carlo simulations. In Fig.
\ref{fig:DiffBetaFM} we plot $k=0$ spectral functions for different
values of $\beta$ for both FM and AFM couplings. In both cases the
$\beta t=5.0$ results are in very good agreement with those at $T=0$,
defining ``low-temperatures'' to mean $\beta t \ge 5$. As $\beta$
decreases ($T$ increases), the sharp peaks broaden considerably and
new peaks appear. For FM coupling, the lowest energy state still lies
at the bottom of the low-energy $T=0$ band. For AFM coupling, however,
new states appear below the $T=0$ spectrum. This is expected since at
finite-$T$ FM domains can form in the AFM background and the
carrier lowers its energy when located in such domains. Of course,
these energies are bounded from below by the lowest FM eigenenergy.

At first sight the appearance of these new peaks (in fact resonances,
as discussed below) may seem to signal lack of convergence of the MC
simulations, or finite-size issues related to a $M_c$ cutoff that is
not big enough. However, the results are converged and do not change
upon further $M_c$ increase; these features are real.

The appearance of similar features is a well documented phenomenon for
the disordered binary alloys with which our problem has similarities, as
discussed above.  Studies of binary alloys have revealed
that these peaks (which are truly discrete states, in that context)
mark the appearance of bound states where the carrier is trapped by
small  clusters of like-atoms.\cite{Economou, Dean} As we show now, the
resonances we observe have similar origin. For instance, in the FM
case they are due to the charge carrier being trapped into  spin-down
domains formed into an otherwise spin-up background, or vice versa.

\begin{figure}[t]
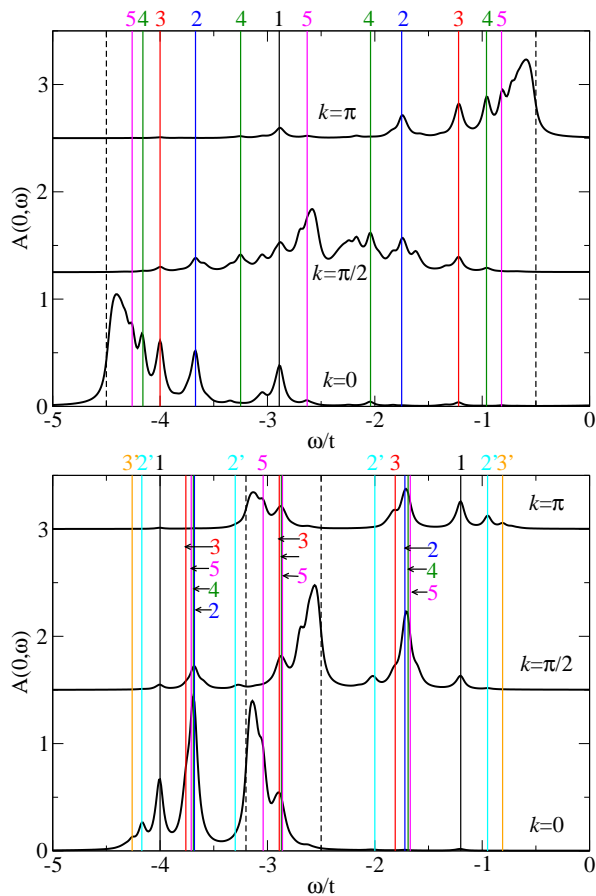

  \includegraphics[width=0.9\columnwidth]{fig3a.eps}
  \includegraphics[width=0.9\columnwidth]{fig3b.eps}
  \caption{(color online) Spectral functions $A(k,\omega)$ for $\beta t =1$ and
  $k=0, \pi/2, \pi$, for FM (top) and AFM (bottom) coupling. Solid vertical
  lines mark the trapping energies of the carrier in small domains. The
  corresponding numbers show the length of the domain (see Table
  \ref{tab:SmallClusters}). The dashed vertical lines mark the band-edges of the
  $T=0$ low-energy band. Other parameters are $|J|/t=0.5,\ J_0/t=2.5,\ h=0,\
  \eta/t=0.04$. }
  \label{fig:SmallClustersFM}
\end{figure}

\begin{table}[b]
    \begin{tabular}{|c|c|c|c|}
      \hline 
      Color & Length & FM domain & AFM domain \\
      \hline
      black &  1 &
      $\dots \uparrow \uparrow \underline{\downarrow} \uparrow \uparrow \dots$ &
      $\dots \uparrow \downarrow \underline{\downarrow} \downarrow \uparrow
      \dots$ \\
      blue  &  2 &
      $\dots \uparrow \uparrow \underline{\downarrow \downarrow} \uparrow
      \uparrow \dots$ &
      $\dots \uparrow \downarrow \underline{\downarrow \uparrow} \uparrow
      \downarrow \dots$ \\ 
      red   &  3 &
      $\dots \uparrow \uparrow \underline{\downarrow \downarrow \downarrow}
      \uparrow \uparrow \dots$ &
      $\dots \downarrow \uparrow \underline{\uparrow \downarrow \uparrow}
      \uparrow \downarrow \dots$ \\ 
      green &  4 &
      $\dots \uparrow \uparrow \underline{\downarrow \downarrow \downarrow
      \downarrow} \uparrow \uparrow \dots$ &
      $\dots \downarrow \uparrow \underline{\uparrow \downarrow \uparrow
      \downarrow} \downarrow \uparrow \dots$ \\
      magenta & 5 &
      $\dots \uparrow \uparrow \underline{\downarrow \downarrow \downarrow
      \downarrow \downarrow} \uparrow \uparrow \dots$ &
      $\dots \uparrow \downarrow \underline{\downarrow \uparrow \downarrow
      \uparrow \downarrow} \downarrow \uparrow \dots $ \\
      cyan &  2' &  - &
      $\dots \uparrow \downarrow \underline{\downarrow \downarrow} \downarrow
      \uparrow \dots$ \\
      yellow &  3' & - &
      $\dots \uparrow \downarrow \underline{\downarrow \downarrow \downarrow}
      \downarrow \uparrow \dots$ \\
      \hline
    \end{tabular}
    \caption{List of the shortest domains (underlined spins) that form in
    otherwise ordered backgrounds. The energies for trapping the carrier in
    these domains are shown in Fig. \ref{fig:SmallClustersFM}.}
    \label{tab:SmallClusters}
\end{table}

We calculated the eigenenergies for trapping the carrier inside
several such short domains embedded in an otherwise ordered FM or AFM
background. These values (various lines) are compared to the spectral
weights obtained for $\beta t=1$ and $k=0,\pi/2,\pi$ in Fig.
\ref{fig:SmallClustersFM}. The integers labelling the lines show the
length of the corresponding domains, also see Table
\ref{tab:SmallClusters}. The agreement between these trapping
energies and the location of the resonances in $A(k,\omega)$ is very
good. The weights of these resonances vary strongly with $k$ but their
energies are nearly dispersionless. For AFM coupling, the trapping
energies in different domains are sometimes very similar, suggesting
that here trapping occurs at the boundaries of the domain, not inside
its bulk; this explains the broader features at $\omega/t \approx
-3.7,-2.9$ and $-1.7$.

\begin{figure}[t]
 \includegraphics[width=0.95\columnwidth]{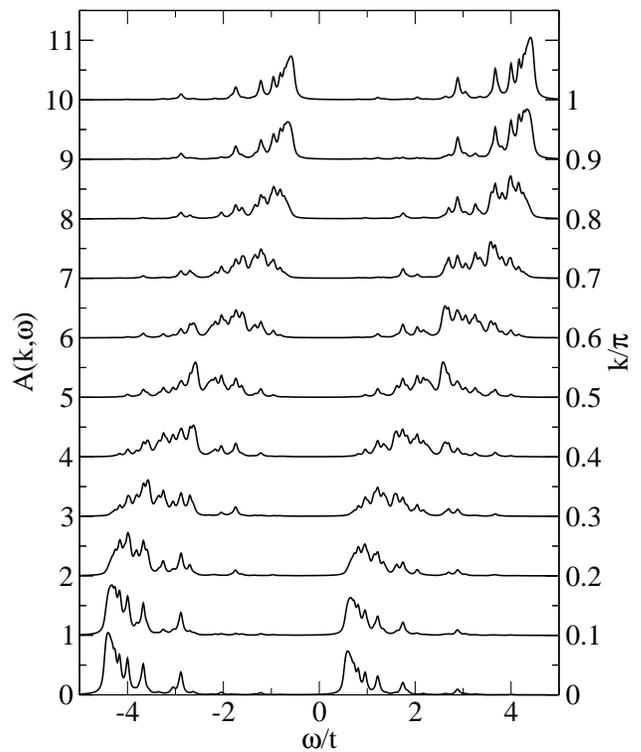}
  \caption{ $A(k,\omega)$ for different values of $k$ and FM coupling, at
  $J/t=0.5,\  J_0/t=2.5,\  \beta t=1.0,\ h=0,\  \eta/ t =0.04$.}
  \label{fig:Diffk}
\end{figure}

\begin{figure}[t]
 \includegraphics[width=0.95\columnwidth]{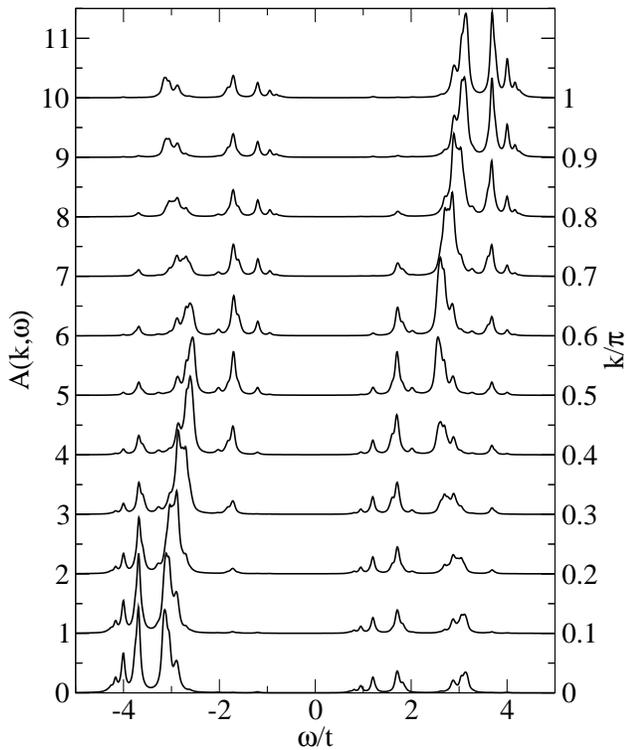}
    \caption{$A(k,\omega)$ for different values of $k$ and AFM coupling, at
    $J/t=-0.5,\  J_0/t=2.5,\  \beta t=1.0,\ h=0,\  \eta/ t =0.04$.}
  \label{fig:DiffkAFM}
\end{figure}

To better understand the momentum dependence, complete sets of
spectral weights are shown in Figs. \ref{fig:Diffk} and
\ref{fig:DiffkAFM} for FM and AFM coupling, respectively. For
simplicity we do not mark the trapping energies of the carrier in the
various domains, but we have checked that the agreement is as good as
in Fig. \ref{fig:SmallClustersFM} at all $k$.

For FM coupling, we see the two bands that have evolved from the $T=0$
peaks moving with increasing $k$ in a way that roughly mirrors the
$T=0$ dispersions shown in Fig. \ref{fig:ZeroTFM}(a). As $k$
increases, spectral weight is systematically shifted from the lower to
the upper edge of each band. In addition we notice a small spectral
weight transfer from the lower band to the upper band. This is in
contrast to the FM $T=0$ solution, where both peaks have equal
spectral weight. For AFM coupling, the $\pi$-periodicity of the $T=0$
dispersion is partially masked by the many additional resonances that
appear on both sides of the $T=0$ peaks, and the
significant transfer of spectral weight from the lower to the upper
band. The latter is similar to the behavior observed for the $T=0$
solution. In both cases, the location of the various resonances does
not change appreciably with $k$.

To understand the physical origin of these resonances, consider the
analogy with the binary alloy model, which also shows such ``peaky''
structures in its total density of states (DOS), marking the bound
states of the carrier inside small clusters of
like-atoms.\cite{Economou, Dean} As is well known, in the presence of
any amount of on-site disorder all eigenstates of a 
1D chain become localized. To find the DOS one can formally calculate the
disorder averaged Green's function (which regains invariance to
translations) but this quantity has no physical meaning. This is
because in any real system there is a given disorder distribution, and
if all eigenstates are localized then the carrier occupies forever the
same small region of space and self-averaging does not occur. In other 
words, if the carrier is trapped in a cluster of atoms it will stay
trapped indefinitely.

Each configuration of the binary alloy can be mapped into a spin
configuration of the Ising chain, by replacing atoms A/B by spins
up/down. Small clusters of like-atoms then map into magnetic domains,
and there are trapped states of the carrier inside them, as already
shown. However, unlike the fixed disorder configuration, the spin
configuration changes continuously through thermal fluctuations. A
trapped carrier therefore has a finite lifetime linked to the
persistence of that domain: eventually the local spin configuration
changes and the carrier is released to move along the chain. This is
why here the ``peaky'' structures indicate actual finite life-time
resonances for extended eigenstates of well-defined momentum, not
infinitely-lived localized states like in the alloy model. This is a
significant qualitative difference.

\begin{figure}[t]
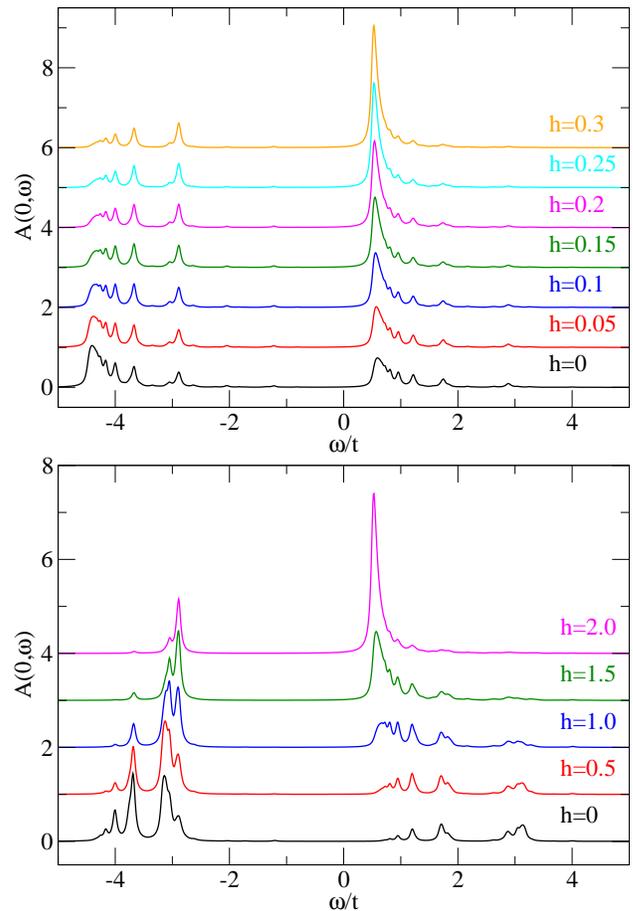

\includegraphics[width=0.95\columnwidth]{fig6a.eps}
\includegraphics[width=0.95\columnwidth]{fig6b.eps}
  \caption{(color online) Spectral functions $A(0,\omega)$ for different values
  of $h$, for FM (top) and AFM (bottom) coupling. Other parameters are
  $|J|/t=0.5,\ J_0/t=2.5,\ \beta t=1.0,\ \eta/ t =0.04$.} 
  \label{fig:DiffhFM}
\end{figure}

We conclude this section by considering the role of the external
magnetic field $h$. Its effects on the $k=0$ spectral weight are shown
in Fig. \ref{fig:DiffhFM}. For FM coupling we find that as $h$
increases, the initially large feature at the bottom of the lower 
band starts to disappear and most of its weight is moved to the bottom
of the upper band. This is expected since the external field
favors/disfavors the spin-up/down background responsible for this main
feature (evolved from the $T=0$ peak) of the
upper/lower band. Interesting, however, most of the resonances in the
lower band remain almost unchanged. The reason is that while $\beta h
\ll 1$, the energy cost for a small domain is very low so their
appearance is very likely. If $\beta h \gg 1$\ the Ising chain is
forced into the $m>0$ ground state and only the higher $T=0$ peak
survives (not shown).

For AFM coupling the effect of $h$ is more dramatic, as it forces the
system into the $m>0$ FM ground-state if $\beta h $ becomes large
enough. Indeed, for large $h$ most of the weight is moved into the
upper FM peak and most of the resonances disappear, except for the one
at $\omega/t \approx -2.9$ that is still quite large. Its energy is
very close to that of the first FM cluster listed in Table
\ref{tab:SmallClusters}, and indeed it seems plausible that this is
due to a single flipped spin which entraps the carrier. This domain is
disfavored by $h$, but actually lowers the exchange energy with its
neighbors.

Note that we used different field strengths for the FM and AFM cases. In
the latter case much higher fields are needed to produce long range
correlations since there is a competition between the exchange energy of
neighboring spins and the external field. A measure for this is the
spin-spin correlation function $\langle
\sigma_{-M_c}\sigma_{M_c}\rangle$, which for the parameters used in
Fig. \ref{fig:DiffhFM} equals 0.41 (FM, if $h/t=0.3$) and 0.64 (AFM,
if $h/t=2.0$). A value of $\langle \sigma_{-M_c} \sigma_{M_c}\rangle
=1$ means that the chain is completely ordered.

\subsection{Analytic approximation}

We now use the insights gained from the Monte Carlo results to propose
an analytic approximation for the GF at low and medium
temperatures. We present the derivation only for the case of FM
coupling when $h=0$; the other cases can be treated similarly.

The main idea is to only consider a limited number of spin
configurations when performing the thermal average, to allow for its
(quasi)analytic evaluation. Since our numerical results show the importance
of small domains, the configurations we select are the two ordered
configurations $| \text{FM}, \sigma \rangle$ with all spins pointing
up or down, $\sigma=\uparrow, \downarrow$ together with the one-domain
configurations:
\begin{align}
  &| n,n+l, \uparrow \rangle = 4^{-(l+1)}
  \hat{\sigma}_n^-\cdot\hat{\sigma}_{n+1}^-\cdot \dots
  \hat{\sigma}_{n+l}^-|\text{FM},\uparrow\rangle \\ &| n,n+l, \downarrow \rangle
  = 4^{-(l+1)} \hat{\sigma}_n^+\cdot\hat{\sigma}_{n+1}^+\cdot \dots
  \hat{\sigma}_{n+l}^+|\text{FM},\downarrow\rangle,
\end{align}
where the domain starts at site $n$ and ends at $n+l$. The operator
$\hat{\sigma}_i^\pm$\ is the raising/lowering operator for the $i^{th}$\ Ising
spin. To preserve translational invariance we need to consider all possible
locations of the domain within the Ising chain. All these one-domain states are
weighed by the same Boltzmann factor $\exponential{-4 \beta J}$ (we take the
energy of the fully ordered FM states as reference).

As discussed, a physically meaningful result has equal contributions
from the spin-down and spin-up sectors. We now discuss the spin-up
contribution, which we denote by $G_{\uparrow}(k,\omega)$. The
spin-down contribution $G_{\downarrow}(k,\omega)$ is then obtained by
simply letting $J_0 \rightarrow -J_0$, and the GF is given by
$G(k,\omega) =  [ G_{\uparrow}(k,\omega) +
G_{\downarrow}(k,\omega)]/2$. By itself, the decomposition into an
up-part and a down-part is not an approximation. The approximation
stems from the fact that we are only considering the one-domain
configurations when calculating $G_{\uparrow}(k,\omega),
G_{\downarrow}(k,\omega)$. 

By only considering domains up to a maximal length $L$ (for reasons
discussed below), we thus approximate:
\begin{align}
  G_{\uparrow}(k,\omega) = \frac{1}{Z} \left [
    G_{\uparrow}^{\text{FM}}(k,\omega) + \exponential{- 4 \beta J}
    \sum_{l=0}^{L} \mathcal{G}_{\uparrow}^{(l+1)}(k,\omega)+\dots
    \right]
\label{app}
\end{align}
where $Z= 1+\exponential{-4\beta |J|} L \cdot N + \dots$. The
thermodynamic limit $N\rightarrow \infty$ will be taken at a later
stage.  The first contribution, from the ordered state, is
\begin{align}
  G_{\uparrow}^{\text{FM}}(k,\omega) = \langle \text{FM},\uparrow | c_{k}
  \hat{G}(\omega+\hat{H}_{\text{I}}) c_{k}^\dagger| \text{FM}, \uparrow \rangle
  \nonumber \\ =\frac{1}{\omega-\varepsilon(k)-J_0+i\eta}.
\end{align}
To find the contributions
\begin{align}
  \mathcal{G}_{\uparrow}^{(l+1)} (k,\omega) = \sum_{n}^{}\langle
  n,n+l,\uparrow | c_{k} \hat{G}(\omega+\hat{H}_{\text{I}})
  c_{k}^\dagger| n,n+l, \uparrow \rangle
\end{align}
 from the states with a domain of length $l$, we have to work
 harder. Using Dyson's identity once we obtain:
\begin{align}
  \mathcal{G}_{\uparrow}^{(l+1)} (k,\omega) =
  G_{\uparrow}^{\text{FM}}(k,\omega) \left [N -
  2J_0\sum_{m=0}^{l}f_{\uparrow,k}^{(l+1)}(m,\omega) \right ],
  \label{eq:GDomain}
\end{align}
where we defined the auxiliary GFs:
\begin{align}
  f_{\uparrow,k}^{(l+1)}(m,\omega) &= \sum_{n}^{}\frac{\exponential{i
  kR_{n+m}}}{\sqrt{N}} \nonumber \\ &\langle n,n+l, \uparrow | c_k
  \hat{G}(\omega+\hat{H}_{\text{I}}) c_{n+m}^\dagger | n,n+l,\uparrow \rangle.
\end{align}
Using
Dyson's equation again we find:
\begin{align}
  f_{\uparrow,k}^{(l+1)}(m,\omega) = G_{\uparrow}^{\text{FM}}(k,\omega)&\\-2 J_0
  \sum_{m'=0}^{l}g_{\uparrow}^{\text{FM}}(m'-m,\omega) \nonumber  
  &\exponential{i k (R_m-R_m')} f_{\uparrow,k}^{(l+1)}(m',\omega),
\end{align}
where $g_{\uparrow}^{\text{FM}}(m'-m,\omega) = {1\over N}\sum_{q}
\exponential{i q (R_{m'}-R_m)} 
G_{\uparrow}^{\text{FM}}(q,\omega)$ are easy to find
analytically. 
This is a linear system of $l+1$ equations that is solved to find all
$ f_{\uparrow,k}^{(l+1)}(m,\omega)$, which are then used in
Eq. (\ref{eq:GDomain}). Note that all
$f_{\uparrow,k}^{(l+1)}(m,\omega)$ are proportional to
$G_{\uparrow}^{\text{FM}}(k,\omega)$, since the latter quantity
provides the inhomogeneous terms in this linear system.

When Eq. (\ref{eq:GDomain}) is inserted in Eq. (\ref{app}), if we
group all terms proportional to $G_{\uparrow}^{\text{FM}}(k,\omega)$
we see that its factor $1/Z$ is cancelled. Higher order terms
corresponding to states with two or more domains (not included in this
calculation) should similarly cancel the factor $1/Z$ for the
remaining terms in Eq. (\ref{app}), or at least make the thermodynamic
limit of the ratio meaningful. To ${\cal O}(e^{-4\beta J})$ order and
for $N\rightarrow \infty$, we
therefore find:
\begin{align}
  G_{\uparrow}(k,\omega) &\approx \left [ 1 - 2 J_0 \exponential{- 4
      \beta J} \sum_{l=0}^{L} \sum_{m=0}^{l}
    f_{\uparrow,k}^{(l+1)}(m,\omega) \right ] \nonumber
  \\ &G_{\uparrow}^{\text{FM}}(k,\omega).
  \label{eq:Approx}
\end{align}
Eq. (\ref{eq:Approx}) obeys the sum rule $\int_{-\infty}^{+\infty}
\mathrm{d}\omega \, A(k,\omega) = 1$ if the second term has no poles
in the upper half of the complex plane. This is because the second
term is proportional to $[G^{\text{FM}}_{\uparrow}(k,\omega)]^2$ and
therefore falls of like $1/\omega^2$ as $|\omega| \rightarrow \infty$.

One may use Eq. (\ref{eq:Approx}) to extract a low-$T$ approximation
for the self energy:
\begin{align}
\Sigma_{\uparrow}(k,\omega)\approx - 2 J_0 \exponential{- 4 \beta J}
\sum_{l=0}^{L} \sum_{m=0}^{l}\frac{
f_{\uparrow,k}^{(l+1)}(m,\omega)}{G_{\uparrow}^{\text{FM}}(k,\omega)},
\end{align}
and define $G_{\uparrow}(k,\omega) \approx \left[
  [G_{\uparrow}^{\text{FM}}(k,\omega)]^{-1}-
  \Sigma_{\uparrow}(k,\omega)\right]^{-1}$ instead of
$G_{\uparrow}(k,\omega)\approx
G_{\uparrow}^{\text{FM}}(k,\omega)\left[1+
  \Sigma_{\uparrow}(k,\omega)G_{\uparrow}^{\text{FM}}(k,\omega)\right]$
of Eq. (\ref{eq:Approx}). At low enough temperatures both give the
same results, but at higher temperatures the former approximation
leads to spurious poles\cite{low-T} in the spectral weight so we use
Eq. (\ref{eq:Approx}) in the following.

\begin{figure}[t]
\includegraphics[width=0.95\columnwidth]{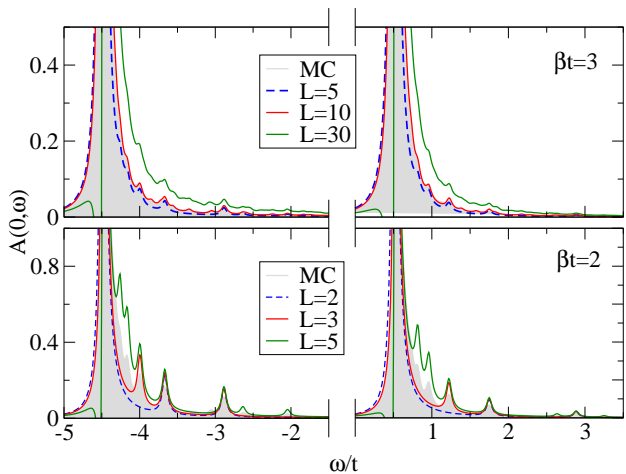}
\caption{(color online) Comparison between the Monte Carlo (MC) results (shaded
area) and the analytic approximation of Eq. (\ref{eq:Approx}) for domains with a
maximal length of $L$, for FM coupling $J/t=0.5$, at $J_0/t=2.5,\ h=0, \eta/t =
0.04,\ \ k=0$ and $\beta t =3$\ (top) and $\beta t =2$\ (bottom).}
  \label{fig:FMDomains}
\end{figure}

\begin{figure}[t]
\includegraphics[width=0.95\columnwidth]{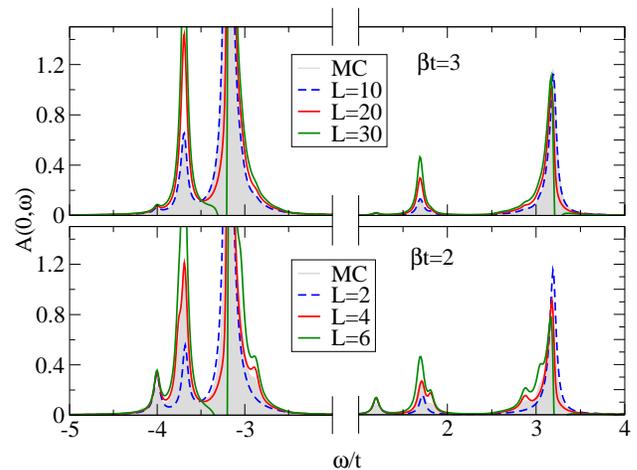}
   \caption{(color online) Same as in Fig. \ref{fig:FMDomains}
     but for AFM coupling 
    $J/t=-0.5$.}
  \label{fig:AFMDomains}
\end{figure}

Results for FM and AFM coupling are shown in Figs. \ref{fig:FMDomains}
and \ref{fig:AFMDomains}, respectively, for various lengths $L$ of the
largest domain included, at two temperatures. For comparison, the
Monte Carlo results are also shown (shaded regions). The quality of
the approximation varies substantially with $L$. The top panel of
Fig. \ref{fig:FMDomains} shows that for $\beta t =3$ and $L = 5$, the
weight of the resonances is underestimated and not all of them are
reproduced by the approximation. For $L=10$ the agreement between the
approximation and the exact results is very good, but it
worsens again for $L= 30$. Not only does the latter overestimate the
weight of the resonances, but it also predicts negative 
spectral weight just below the bands. While this negative weight is
needed to satisfy the sum rule, its presence is unphysical and signals
a failure of the approximation.  
The same trends are observed for $\beta t =2$ in the bottom panel of
Fig. \ref{fig:FMDomains}. Here the best agreement is obtained at
$L=3$, although resonances associated with longer domains are
missing. They appear for $L=5$, however so does the unphysical
behavior. For even lower values of $\beta$ the approximation fails
completely to capture the correct weight of the resonances,
although, as shown in Fig. \ref{fig:SmallClustersFM}, their locations
are due to carrier trapping in domain walls.

The AFM approximation yields very similar results. The top panel of
Fig. \ref{fig:AFMDomains} shows that again for $\beta t = 3$
excellent agreement with the exact solution is reached for $L=10$,
while for larger values of $L$ the weight of the resonances is
overestimated and unphysical behavior occurs if $L=30$. For $\beta t =
2$ the agreement with the exact solution is best for $L=4$ and
unphysical behavior already occurs at $L=6$. Again the approximation
fails badly to capture the proper weight of various features, for
smaller values of $\beta$ (higher $T$).

Naively, one may expect the approximation to improve when $L$ is
increased since this means that a larger fraction of the possible
configurations is considered. However, there are two factors which
determine how a domain contributes to the thermal average. One is the
additional energy cost of a domain, which is accounted for by the
Boltzmann factor and in 1D does not depend on the domain's length.
The other is the increase in entropy with increasing number of
domains. As the temperature increases, minimization of the free energy
$F=U-TS$ is increasingly driven by entropy maximization, resulting in
more domains and thus shorter correlations. The order of magnitude for
the maximal domain size should be given by the spin-spin correlation
length $\xi$ (defined in the Appendix). Indeed, we obtain $\xi\approx4$ and $\xi\approx10$ for 
$\beta t = 2 $ and $\beta t = 3$, respectively. This compares well
with the values of $L$ where the approximation performs well, see
Figs. \ref{fig:FMDomains} and \ref{fig:AFMDomains}.

Another way to see why the approximation with $L\rightarrow \infty$ is
bound to become wrong is to realize that all domain walls whose length
is longer than the distance explored by the carrier within its
lifetime are actually indistinguishable from the ``other'' ordered FM
configuration, from the point of view of the carrier. In other words,
all these configurations essentially contribute a
$G_{\downarrow}^{\text{FM}}(k,\omega)$, and their inclusion gives the
wrong weighting to the $|\text{FM}, \downarrow\rangle$
contribution. Similarly, configurations with two long domain walls
placed relatively close together will have states where the carrier is
trapped in the short region between the domains, indistinguishable from
having a short domain formed in the ``other'' ordered FM
configuration. Adding many such contributions will affect the weights
of these one-domain contributions, etc. These arguments suggest that a
better approximation is:
\begin{align}
  G_{\uparrow}(k,\omega) = \left [ 1 - \sum_{l=0}^{N/2}
    w_l(\beta)\sum_{m=0}^{l} f_{\uparrow,k}^{(l+1)}(m,\omega) \right
  ]G_{\uparrow}^{\text{FM}}(k,\omega)
\end{align}
with $\left. G_{\downarrow}(k,\omega)
=G_{\uparrow}(k,\omega)\right|_{J_0\rightarrow -J_0}$, where $
w_l(\beta)$ are adjusted to capture accurately the weight of
resonances due to trapping into short domains. For low and medium
temperatures we showed that $w_l(\beta)= 2J_0e^{-4\beta J}$ if $l \le
\xi$, and zero otherwise, gives very decent predictions. Clearly this
cannot work at high temperatures of order $\beta t =1$ where $\xi
\rightarrow 0$. So far we have been unable to think of a reasonable
form of $w_l(\beta)$ in this regime, but the comparisons displayed in
Fig. \ref{fig:SmallClustersFM} suggest that it should exist.

\section{Conclusions}

To summarize, we obtained numerically exact spectral functions for a simplified
model of a carrier injected into a 1D Ising chain at finite-$T$. The
results highlight the importance of small domains that can trap the
carrier, which were shown to be responsible for the resonances
that appear as $T$ increases. A simple
analytic approximation based on these ideas was found to perform
well at low and medium temperatures. With further insights, it may be
possible to generalize it to high temperatures, as well.
Interestingly, chains with both FM and AFM coupling can be understood
in similar terms, although generically one expects quite different
phenomenology for a carrier injected into a FM vs AFM background.

As highlighted throughout, there are parallels between this problem
and that of a carrier moving in a random binary alloy, where the
importance of small clusters of like-atoms, equivalent to the small
domains of our model, is well documented.\cite{Dean, Economou, Alben}
There are, however, also major differences: finite $J$ maps into
correlations between the atoms of the alloy (usually these are
ignored). The thermal average is also very different, both
qualitatively and quantitatively, from a disorder average. It is
therefore not a priori clear how much of the considerable amount of
work devoted to finding analytic approximations for binary alloys
can be used for the magnetic problem.

In terms of generalizations, one direction is to see how far these
insights carry over to higher dimensions, where long-range magnetic
order survives at finite $T$. For binary alloys it is known that the
fine-structure of the spectral function is most dominant in
1D.\cite{Economou} For our model, the energy cost scales with the
domain size in 2D and 3D, unlike in 1D and unlike for a binary
alloy. This may well lead to a behavior that is different from that of
the binary alloys and could cause the fine-structure associated with
very small domains to remain present in higher dimensions, at least at
low temperatures.

Another interesting direction is to allow for Heisenberg coupling between the
carrier and the local moments (still with Ising coupling between the latter).
In this case spin-flip processes become possible. The carrier can move small
domains around (in fact, the spin-polaron\cite{Shastry,Mona-sp} can be thought
of as a mobile bound state between the carrier and a one-site domain), or split
longer domains into several smaller ones, etc. Understanding the consequences of
such processes and their effect on the finite-$T$ spectral function would be
very useful. Even more complicated is the case with Heisenberg coupling between
the lattice spins. Clearly there is a lot of work, both numerical and
analytical, to be done before these problems are solved.

Our results underline the importance of the local environment for the
behavior of a charge carrier in a magnetic background, at least for
this model and in low dimension. Incorporating
these effects is difficult since mean-field approximations cannot
capture them. The only real alternative is to obtain an understanding
of which states of the environment contribute most to the temperature
average, and to propose approximations based on taking the average
over this limited set of states. Our work presents the first step in
this process.

\appendix*
\section{Exact solution of the undoped Ising chain}
Here we review the exact solution of the undoped Ising chain.
All quantities
of interest to us are obtained from the partition function:
\begin{align}
  Z = \sum_{\{\sigma\}}\exponential{-\beta E_I(\{\sigma\})},
\end{align}
where the sum is over all configurations of lattice spins.
The sum 
can be rewritten as:
\begin{align}
  Z  = \tr (\mathcal{T}^N) \nonumber 
  = \lambda_+^N + \lambda_-^N,
\end{align}
where the transfer matrix is $\mathcal{T}_{\sigma,\sigma'}
=\exponential{\beta (J \sigma \sigma' +\frac{h}{2}(\sigma + \sigma'))
}$, and its
eigenvalues are: 
\begin{align}
  \lambda_{\pm} = \exponential{\beta J} \left [\cosh(\beta h) \pm
  \sqrt{\sinh^2(\beta h) + \exponential{-4 \beta J}} \right].
\end{align}
The bulk value of the magnetization $m={1\over N} \sum_{i}^{} \langle
\hat{\sigma}_i\rangle_T$ is: 
\begin{align}
  m = \lim_{N\rightarrow \infty} {1\over N \beta} \frac{\partial \ln Z}{\partial h}=
\frac{\sinh(\beta h)}{\sqrt{\sinh^2(\beta h) + 
  \exponential{-4 \beta J}}}.
\end{align}
The correlation between spins is given by
\begin{align}
  \langle \sigma_n \sigma_m \rangle_T =
  \frac{\sinh^2(\beta h) + \exponential{-4 \beta J} 
  \exponential{-| m-n |/ \xi}
  }{\sinh^2(\beta h) + \exponential{-4 \beta J}},
\end{align}
where the correlation length is $\xi= 1/
\log(\lambda_-/\lambda_+) $. (For AFM coupling one needs to factor out
$(-1)^{|m-n|}$\ to ensure the real-valuedness of $\xi$).

\acknowledgements
M.M. is grateful to S. Johnston and D. Marchand for help with the Monte Carlo
code. Financial support from NSERC and the UBC Four Year Doctoral Fellowship
program are acknowledged.


%

\end{document}